\documentclass[showpacs, pra,onecolumn,preprintnumbers ,amsmath, amssymb, superscriptaddress, aps]{revtex4-2}
\usepackage{color}
\usepackage{amsmath,amssymb}
\usepackage{pifont}
\usepackage{amssymb}  
\usepackage{bbold}
\usepackage{float}
\usepackage{subfloat}

\usepackage[caption=false]{subfig}
\usepackage{tikz}
\usepackage{makecell}
\usepackage{subfig}
\usepackage{pifont}   
\usepackage{graphicx} 
\graphicspath{{Figuress/}}
\usepackage{dcolumn}  
\usepackage{bm}       
\usepackage{multirow} 
\usepackage{placeins}
\usepackage[colorlinks]{hyperref}
\usepackage{mathtools}
\usepackage{appendix}

\def \be{\begin{align}}
	\def \ee{\end{align}}
\def \bea{\begin{eqnarray}}
	\def \eea{\end{eqnarray}}

\begin{document}
	
	\title{Conductance in graphene through double laser barriers and magnetic field
	}
	
	\author{Rachid El Aitouni}
	\affiliation{Laboratory of Theoretical Physics, Faculty of Sciences, Choua\"ib Doukkali University, PO Box 20, 24000 El Jadida, Morocco}
	
	\author{Miloud Mekkaoui}
	\affiliation{Laboratory of Theoretical Physics, Faculty of Sciences, Choua\"ib Doukkali University, PO Box 20, 24000 El Jadida, Morocco}
	\author{Abdelhadi Bahaoui}
	\affiliation{Laboratory of Theoretical Physics, Faculty of Sciences, Choua\"ib Doukkali University, PO Box 20, 24000 El Jadida, Morocco}
	\author{Ahmed Jellal}
	\email{a.jellal@ucd.ac.ma}
	\affiliation{Laboratory of Theoretical Physics, Faculty of Sciences, Choua\"ib Doukkali University, PO Box 20, 24000 El Jadida, Morocco}
	\affiliation{Canadian Quantum  Research Center,
		204-3002 32 Ave Vernon,  BC V1T 2L7,  Canada}
	\begin{abstract}
	Photon-assisted charge transport through a double barrier laser structure, separated by a region assisted by a magnetic field, is studied. Employing Floquet theory and matrix formalism, the transmission probabilities for the central band and sidebands are calculated. The temporal periodicity of the laser fields creates an infinite number of transmission modes due to the degeneracy of the energy spectrum. The challenge of numerically addressing all modes necessitates the limitation to the first sideband corresponding to energies $\varepsilon\pm\varpi$. A critical phase difference between the two laser fields is found to cancel the transmission through the sidebands due to quantum interference. Varying the width of the region where the magnetic field is applied allows for the suppression of lateral transmission and control over the transmission mode. The intensity of the laser fields also allows for suppressing Klein tunneling and blocking transmission processes with zero photon exchange, as well as activating transmission processes with photon exchange. The conductance is also affected by changes in the system parameters. Increasing the intensity of the laser field reduces the conductance due to the confinement of the fermions by the laser fields. In addition, increasing the size of the region where the magnetic field is applied reduces the conductance because the increased distance gives the fermions a greater chance of diffusion and increases their interaction with the magnetic field.
	\end{abstract}

	\maketitle

\section{Introduction}

Technological development is constantly searching for revolutionary materials to produce electronic components with lower resistance and higher conductivity. While traditional semiconductors such as germanium and silicon played a critical role in the previous cycle, their performance remains modest. In recent years, new materials such as compound semiconductors, for example GaAs, GaSb $\cdots$, have contributed to the evolution of electronic components. In 2004, researchers Geim and Novoselov successfully isolated a graphene sheet for the first time \cite{disc}, paving the way for a new technology based on this material. Due to its extraordinary properties, graphene has a remarkably high conductivity \cite{prop}. Its electrons move at a speed about $300$ times slower than that of light \cite{mobile}, called Dirac fermions \cite{masless}. Graphene is also flexible, much harder than other materials \cite{mobil2}, and transparent, absorbing only about $2.3\%$ of incident light \cite{absor}. Composed entirely of carbon atoms, graphene is one of the rarest two-dimensional materials, with its atoms arranged in hexagonal shapes due to $sp^2$ hybridization \cite{sp2}. Despite these remarkable properties, the use of graphene in the fabrication of electronic devices remains limited due to the absence of a band gap in its energy spectrum \cite{zero, desp2,zero1}. The valence and conduction bands touch at six points, known as Dirac points, making the electrons uncontrollable and impossible to stop \cite{Tight}. Following this discovery, researchers have explored various techniques to create a bandgap to control the passage of fermions between the two bands. They have succeeded in creating this bandgap by applying a potential barrier in a specific region to confine the fermions. However, a new problem has arisen: despite the formation of this bandgap, fermions incident perpendicular to the barrier pass through it completely without reflection, a phenomenon called Klein tunneling \cite{klien1, klien2, klienexp}.

One of the most fascinating areas of research at the moment is the confinement of Dirac fermions in graphene, where new results are published daily. Numerous methods have been discovered, including doping graphene with different types of atoms \cite{dopage}, which facilitates the formation of bandgaps. Creating pseudomagnetic fields by deforming graphene sheets \cite{def1,def2}. The bandgap can also be created by depositing graphene on a substrate \cite{substrat} or by surrounding it with magnetic barriers \cite{mag1,mag4}.
However, the main challenge remains Klein tunneling, although laser irradiation has been somewhat effective in mitigating this phenomenon \cite{Elaitouni2023, Elaitouni2023A}. Magnetic barriers, on the other hand, lead to quantization of the energy spectrum, producing what are known as Landau levels \cite{mag3}. As the intensity of the magnetic field increases, the energy spacing between successive levels decreases. 
Time-oscillating barriers also lead to quantization of the energy spectrum, introducing two transmission processes: one with photon exchange and one without \cite{oscil1, oscil2}. Photon exchange between the field and the fermions plays a role in creating the bandgap. Similarly, laser irradiation creates two transmission paths—one with photon exchange and one without \cite{bis2, doublelaser}. The intensity of the laser serves as a means of controlling the transmission process and can also suppress the Klein tunneling effect.

We study the conductance of a graphene system consisting of five regions with two laser barriers separated by a region containing an applied magnetic field.
After obtaining the energy spectrum solutions, we use the boundary conditions together with the transfer matrix approach to determine the transmission coefficients. This is because the energy spectrum is quantified by the temporal oscillation of the fields, resulting in an infinite number of transmission modes.
We then explicitly derive the transmission  from the current density and subsequently the corresponding conductance. 
For a better understanding, we analyze our results numerically. Due to computational limitations, we focus only on the first three channels: one with photon exchange between the barrier and the fermions, corresponding to the sidebands, and another without photon exchange, corresponding to the central band. 
We show that increasing the laser field intensity allows the inhibition of transmission without photon exchange and initiates the process with photon exchange. Additionally, we find that the total transmission is reduced by increasing the region over which the magnetic field is applied. At normal incidence, the Klein tunneling occurs, but it can be inhibited by increasing the laser field intensity.
On the other hand, based on the obtained transmissions, we determine the associated conductance. We show that as the laser field intensity increases, the conductance decreases. Furthermore, it is found that the distance separating the two barriers—the region of the applied magnetic field—affects the conductance. We conclude that this distance should be kept small for better conductance.


The manuscript is organized as follows: in Sec. \ref{TM}, we present the theoretical model describing the movement of an electron through our structure. In Sec. \ref{TP}, we determine the transmission coefficients for each energy band using the transfer matrix method and the current density in each region. In Sec. \ref{NR} we present our numerical results as a function of various parameters of our system. Finally, we conclude in Sec. \ref{CC}.

\section{THEORETICAL MODEl}\label{TM}
\begin{figure}[H]
	\centering
	\includegraphics[scale=0.35]{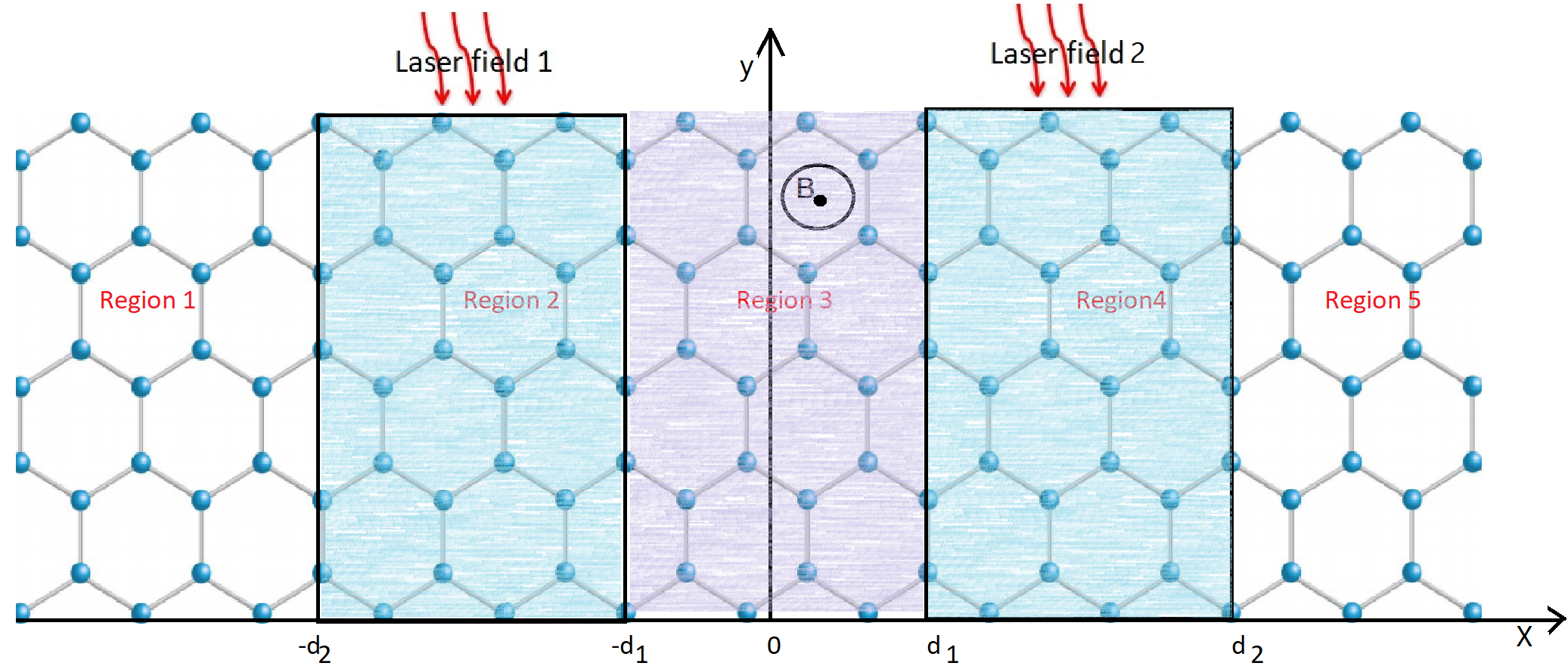}
	\caption{(Color online) Schematic of a graphene sheet divided into 5 regions, where regions 2 and 4 are irradiated by two laser fields, while region 3 is subjected to a magnetic field perpendicular to the sheet.}\label{fig1}
\end{figure}

We apply two laser fields generated by two electric fields in the electric dipole approximation and a magnetic field perpendicular to the graphene sheet. The three fields divide the sheet into five regions indexed by $j=1, \cdots, 5$. In regions 1 and 5, we only have pristine graphene. In regions 2 and 3, we apply the two laser fields with different amplitudes and phase-shifted by $\beta$. In region 3, we have the magnetic field, as shown in Fig. \ref{fig1}.
In the presence of the  two laser barriers and magnetic field,
the Hamiltonian describing the present system can be written as
\begin{equation}\label{H1}
		H=v_F\vec{\sigma}_x \vec{p}_x+v_F\vec{\sigma}_y \left(\vec{p}_y+\frac{e}{c}{\vec{A}_{j}}(t)+{\vec{A}_B}(x)\right)
	\end{equation}
where $v_{F}$ is the Fermi velocity, $\vec{\sigma}=\left(\vec{\sigma}_{x}, \vec{\sigma}_{y}\right)$ are the Pauli matrices. The vector potential $\vec{A}_{lj}$ corresponding to the laser field of frequency $\omega$, in the dipole approximation \cite{dipole}, is given by 
%
%
\begin{align}	{A}_j(t)=
	\begin{cases}
		A_2\cos(\omega t),&-d_2<x<-d_1\\
		A_4\cos(\omega t+\beta),&d_1<x<d_2\\
		0,& \text{otherwise}
	\end{cases}
\end{align}
where $\beta $ is a phase shift. Note that the interface positions of the two laser barriers are identified as ($-d_2,-d_1$) for the first and ($d_1,d_2$) for the second. The vector potential $\vec{A}_B(x)$ corresponding to the magnetic field in Landau gauge takes the form
\begin{equation}A_{B}(x)=\frac{1}{l_{B}^{2}} 
\begin{cases}
	-d_{1}, & x<d_{1} \\
	 x, & |x|<d_{1} \\ 
	 d_{1}, & x>d_{1}\end{cases}
\end{equation}
and  in the unit system $(\hbar=c=e=1)$, the magnetic length is $l_{B}=\sqrt{\frac{1}{B}}$.

To find the eigenspinors in each region, we solve the wave equation by separating the Hamiltonian \eqref{H1} into spatial and temporal components, denoted as \( H_0 \) and \( \widetilde{H}_j \), respectively. Specifically, we express the total Hamiltonian as \( H_j = H_0 + \widetilde{H}_j \), where \( H_0 \) accounts for the spatial part and \( \widetilde{H}_j \) represents the time-dependent part, as follows
	\begin{align}
	&	H_0=v_F  \left(\sigma_x p_x+\sigma_y (p_y+{A_B}(x))\right)\\
	& \widetilde{H}_j=v_F \sigma_y A_j(t).
	\end{align}
	Given that $H_j$ is time-independent and $\widetilde{H}_j$ is coordinate-independent, the total eigenspinors $	\Psi_j(x,y,t)=\phi_j(x,y)\chi_j(t) $ can be represented as a tensor product of the two eigenvectors  {$\psi_j(x,y)=\dbinom{\phi_{j}^+(x,y)}{\phi_{j}^-(x,y)}$} and $\phi_j(t)$ associated with  $H_0$ and $\widetilde{H}_j$, respectively.
	In the framework of the Floquet approximation \cite{bessel,floq}, the temporal part is expressed as $\chi_j(t)=\xi_j(t)e^{-i\varepsilon t}$ 
	associated with the Floquet energy $\varepsilon=\frac{E}{v_F}$, with $\xi_j(t)$ being a periodic function over time. The eigenvalue equation $H_j\Psi_j(x,y,t)=E\Psi_j(x,y,t)$  yields
	\begin{align}
		&\label{eq8}	\left[\partial_x +(k_y+{A_B}(x))-A_j \cos\Phi_j\right] \phi_{j}^-(x,y)\xi_j(t)=\phi_{j}^+(x,y)\frac{\partial}{\partial t}\xi_j(t)\\
		&\label{eq9} 	\left[\partial_x -(k_y+{A_B}(x))+	A_j \cos\Phi_j\right]  \phi_{j}^+(x,y)\xi_j(t)=\phi_{j}^-(x,y)\frac{\partial}{\partial t}\xi_j(t)
	\end{align}
	where $\Phi_j$ is either $\omega t$ or   $\omega t+\beta$.
	Unfortunately, solving the above system is impossible due to the presence of three unknown functions (\(\phi_j^+\), \(\phi_j^-\), and \(\xi_j\)). To deal with this, we can make an approximation \cite{bis2} by assuming that inside the barrier, the laser-free coupled differential equations are satisfied by \(\phi_j^+\) and \(\phi_j^-\). Consequently, \eqref{eq8} and \eqref{eq9} can be simplified as 
	\begin{align}
		&\label{eq88}	-A_j \cos\Phi_j\  \phi_{j}^-(x,y)\xi_j(t)=\phi_{j}^+(x,y)\frac{\partial}{\partial t}\xi_j(t)\\
		&\label{eq99}	A_j \cos\Phi_j\  \phi_{j}^+(x,y)\xi_j(t)=\phi_{j}^-(x,y)\frac{\partial}{\partial t}\xi_j(t)
	\end{align}
	which leads to the following second-order differential equation
	\begin{equation}
		\left({\partial_t^2}+\omega \tan\Phi_j\ \partial_t+A_j^2\cos^2\Phi_j\right) \xi_j(t)=0
	\end{equation}
	and the corresponding solution can be written as
	\begin{equation}
		\xi_j(t)=e^{-i\alpha\sin\Phi_j}=\sum_{m=-\infty}^{+\infty}J_m(\alpha_j)e^{-m\Phi_j}
	\end{equation}
	where $J_m$ is the Bessel functions.
As a result, the oscillation of the laser fields creates several energy bands $\varepsilon+l\varpi$, with $l=0,\pm 1, \pm 2\cdots$ \cite{oscil1,oscil2,timepot,timepot2,doubletemps}, then the eigenspinors   in each region can be written as 
%
%
\begin{equation}
	\Psi_{j}(x,y,t)=\sum_{m,l=-\infty}^{\infty}\phi^l_j(x,y,0)J_{m-l}(\alpha_j)e^{-i(m-l)\beta}e^{-iv_F(\varepsilon+\varpi)t}.
\end{equation}
where we have set the quantities $\varpi=\frac{\omega}{v_F}$, $\varepsilon=\frac{E}{v_F}$, $A_j=\frac{F_j}{\varpi}$ and $\alpha_j=\frac{F_j}{\varpi^2}$. Next, we explicitly determine the eigenspinors for each region.

In regions 1 and 5, where pristine graphene is present, the corresponding spinors can be expressed as \cite{Elaitouni2022,Elaitouni2023}
\begin{align}
&	\Psi_1(x,y,t)=\sum_{l,m=\infty}^{\infty}\left[\begin{pmatrix}
		1\\
		\Xi^l_1
	\end{pmatrix}\delta_{m,0}e^{ik^0_1(x-x_1)}+r_l\begin{pmatrix}
		1\\
		-\frac{1}{\Xi^l_1}
	\end{pmatrix}e^{-ik^l_1(x-x_1)}\right]e^{ik_yy}\delta_{m,l}e^{-iv_F(\varepsilon+m\varpi)t}\\
&	\Psi_5(x,y,t)=\sum_{l,m=\infty}^{\infty}\left[t_l
	\begin{pmatrix}
		1\\
	\Xi^l_5
		\end{pmatrix}e^{ik^l_5(x-x_5)}+\mathbb{0}_l\begin{pmatrix}
		1\\
		-\frac{1}{\Xi^l_5}
	\end{pmatrix}
	e^{-ik^l_5(x-x_5)}\right]e^{ik_yy}\delta_{m,l}e^{-iv_F(\varepsilon+m\varpi)t}\\
&	\Xi^l_{1,5}=s_l\frac{k^l_{1,5}+i\left(k_y\mp\frac{d_1}{l^2_B}\right)}{\sqrt{\left(k^l_{1,5}\right)^2+\left(k_y\mp\frac{d_1}{l^2_B}\right)^2}}=s_le^{i\theta_l},\quad
	\theta^l_{1,5}=\arctan \frac{k_y\mp\frac{d_1}{l^2_B}}{k_{1,5}^l}
\end{align}
associated the the energies
\begin{align}
\varepsilon+l\varpi=s_l\sqrt{\left(k^l_{1,5}\right)^2+\left(k_y\mp\frac{d_1}{l^2_B}\right)^2}	
\end{align}
with the sign function $ s_l=\text{sgn}(\varepsilon+l\varpi)$, $\delta_{m,l}=J_{m-l}(0)$, $\{\mathbb{0}_l\}$ is the null vector, and $c_{il}$ $(i=1,2)$ are two constants. The coefficients $r_l$ and $t_l$ denote the reflection and transmission amplitudes, respectively, which can be determined from the boundary conditions.

In regions $2$ and $4$, where we apply the laser fields, the eigenvalue equation allows us to obtain
\begin{eqnarray}
		\left(-i\partial_x-i\left(k_y\mp\frac{d_1}{l^2_B}-m\varpi\right)\right)\phi_{2j}(x,y)&=&(\varepsilon+m\varpi)\phi_{1j}(x,y)\\
		\left(-i\partial_x+i\left(k_y\mp\frac{d_1}{l^2_B}-m\varpi\right)\right)\phi_{1j}(x,y)&=&(\varepsilon+m\varpi)\phi_{2j}(x,y).
\end{eqnarray}
These can be solved by adopting the approach used in \cite{doublelaser,doubletemps} to end up with the solutions
\begin{align}
&	\Psi_2(x,y,t)=\sum_{l,m=\infty}^{\infty}\left[a^2_{l}\begin{pmatrix}
		1\\
		\chi_2^l
	\end{pmatrix}e^{iq^l_{2}(x-x_2)}+b^2_{l}\begin{pmatrix}
		1\\
		-\frac{1}{\chi^l_2}
	\end{pmatrix}e^{-iq^l_{2}(x-x_2)}\right]e^{ik_yy}J_{m-l}(\alpha_2)e^{-iv_F(\varepsilon+m\varpi)t}\\
&	\Psi_4(x,y,t)=\sum_{l,m=\infty}^{\infty}\left[a^4_{l}\begin{pmatrix}
		1\\
	\chi^l_4
	\end{pmatrix}e^{iq^l_{4}(x-x_4)}+b^4_{l}\begin{pmatrix}
		1\\
		-\frac{1}{\chi^l_4}
	\end{pmatrix}e^{-iq^l_{4}(x-x_4)}\right]e^{ik_yy}J_{m-l}(\alpha_4)e^{-iv_F(\varepsilon+m\varpi)t}e^{-i(m-l)\beta}\\
	&	\chi^l_{2,4}(x)=s_l\frac{q^l_{2,4}+i(k_y\mp\frac{d_1}{l^2_B}-l\varpi)}{\sqrt{(q^l_{2,4})^2+(k_y\mp\frac{d_1}{l^2_B}-l\varpi)^2}}=s_le^{i\theta_l'}, \quad
	\theta_l'=\arctan\frac{k_y\mp\frac{d_1}{l^2_B}-l\varpi}{q^l_{2,4}}
\end{align}
and the  energies 
\begin{align}
\varepsilon+l\varpi=s_l\sqrt{(q^l_{2,4})^2+(k_y\mp\frac{d_1}{l^2_B}-l\varpi)^2}	
\end{align}
where $a(b)^i_{l}$ are constants ($i=2,4$).

In the region $j=3$, where a magnetic field is applied, we introduce the creation $a_l$ and annihilation $a_l^\dagger$ operators to diagonalize the Hamiltonian. They are
\begin{align}
	a_{l}=\frac{l_{B}}{\sqrt{2}}\left(\partial_{x}+k_{y}+\frac{x}{l_{B}^{2}}\right), \quad a_{l}^{\dagger}=\frac{l_{B}}{\sqrt{2}}\left(-\partial_{x}+k_{y}+\frac{x}{l_{B}^{2}}\right)
\end{align}
 which satisfy the commutation relation $\left[a_{l}, a_{k}^{\dagger}\right]=\delta_{l k}$. As a result, the eigenvalue equation implies
\begin{align}
-i \frac{\sqrt{2}}{l_{B}} a_{l} \phi_{l, 2}=\left(\epsilon+l \varpi\right) \phi_{l, 1}\label{coup1}, \quad
 i \frac{\sqrt{2}}{l_{B}} a_{l}^{\dagger} \phi_{l, 1}=\left(\epsilon+l \varpi\right) \phi_{l, 2} 
\end{align}
These can be used to map an equation for $\phi_{l,1}$
%
\begin{equation}
	\left(\epsilon+l \varpi\right)^{2} \phi_{l, 1}=\frac{2}{l_{B}^{2}} a_{l} a_{l}^{\dagger} \phi_{l, 1}
\end{equation}
It is obvious that $\phi_{l,1}$ is proportional to the state $|n-1\rangle$, since it is an eigenstate of the number operator $\hat{N}=a_{l}^{\dagger} a_{l}$.
Moreover, $\phi_{l,1}$ can be expressed in cylindrical parabolic functions, similar to a harmonic oscillator. Consequently, the solution in region 3 can be written as \cite{magneticfield,masse}
\begin{align}
&	\Psi_{3}(x, y, t)=\sum_{l,m=-\infty}^{l=+\infty}\left[a_{l}^{3}\binom{\mu_l^+}
	{	\xi_l^+}
 +b_{l}^{3} \binom{\mu_l^-}
	{\xi_l^-}
\right] e^{i k_{y} y}\delta_{m,l}e^{-i v_{F}(\epsilon+l \varpi) t}
\\
	&\mu_{l}^{ \pm}(x)=D_{n-1}\left[ \pm \sqrt{2}\left(\frac{1}{l_{B}}(x-x_3)+k_{y} l_{B}\right)\right]\\
	&\xi_l^\pm(x)=\pm i \frac{\sqrt{2}}{l_B (\epsilon-l\varpi)} D_{n}\left[ \pm \sqrt{2}\left(\frac{1}{l_{B}}(x-x_3)+k_{y} l_{B}\right)\right]\\
\end{align}
and the corresponding eigenvalues are 
\begin{equation}
	\epsilon+l \varpi= \pm \frac{1}{l_{B}} \sqrt{2 n}
\end{equation}
with $n$ is an integer value.
We will explore how the above findings can be applied to study the transport properties of the current system in the upcoming investigation.

\section{Transmission probability}\label{TP}
We determine the transmission coefficients by applying the boundary conditions to the eigenvectors. This process yields eight equations, each involving an infinite number of modes, taking into account the orthogonalization $e^{iv_F m \varpi t}$. Then, at interfaces  $x = -d_2,-d_1,d_1,d_2$, we have the continuities, respectively, $ \Psi_1(-d_2,y,t) = \Psi_2(-d_2,y,t), \Psi_2(-d_1,y,t)=\Psi_3(-d_1,y,t), \Psi_3(d_1,y,t)=\Psi_4(d_1,y,t), \Psi_4(d_2,y,t)=\Psi_5(d_2,y,t)$
giving raise to the following set of equations
\begin{align}
&\delta_{m,0}+r_m=\sum_{l=-\infty}^{\infty}\left[a^2_l+b^2_l\right]J_{m-l}(\alpha_2)\\
&\delta_{m,0}\Xi^m_1-r_m\frac{1}{\Xi^m_1}=\sum_{l=-\infty}^{\infty}\left[a^2_l\chi^l_2-b^2_l\frac{1}{\chi^l_2}\right]J_{m-l}(\alpha_2)\\
	&a^3_m\mu^+_m(-d_1)+b^3_m\mu_m^-(-d_1)=\sum_{l=-\infty}^{\infty}\left[a^2_le^{iq^l_{2}(d_2-d_1)}+b^2_le^{-iq^l_{2}(d_2-d_1)}\right]J_{m-l}(\alpha_2)\\
	&a^3_m\xi^+_m(-d_1)+b^3_m\xi_m^-(-d_1)=\sum_{l=-\infty}^{\infty}\left[a^2_l\chi^l_2 e^{iq^l_{2}(d_2-d_1)}-b^2_l\frac{1}{\chi^l_2}e^{-iq^l_{2}(d_2-d_1)}\right]J_{m-l}(\alpha_2)\\
	&a^3_m\mu^+_m(d_1)+b^3_m\mu_m^-(d_1)=\sum_{l=-\infty}^{\infty}\left[a^4_l+b^4_l\right]J_{m-l}(\alpha_4)e^{-i(m-l)\beta}\\
	&a^3_m\xi^+_m(d_1)+b^3_m\xi_m^-(d_1)=\sum_{l=-\infty}^{\infty}\left[a^4_l\chi^l_4-b^4_l\frac{1}{\chi^l_4}\right]J_{m-l}(\alpha_4)e^{-i(m-l)\beta}\\
	&t_m+\mathbb{0}_m=\sum_{l=-\infty}^{\infty}\left[a^4_le^{iq^l_{4}(d_2-d_1)}+b^2_le^{-iq^l_{4}(d_2-d_1)}\right]J_{m-l}(\alpha_4)e^{-i(m-l)\beta}\\
	&t_m\Xi^m_5-\mathbb{0}_m\frac{1}{\Xi^m_5}=\sum_{l=-\infty}^{\infty}\left[a^4_l\chi^l_4 e^{iq^l_{4}(d_2-d_1)}-b^4_l\frac{1}{\chi^l_4}e^{-iq^l_{4}(d_2-d_1)}\right]J_{m-l}(\alpha_4)e^{-i(m-l)\beta}
\end{align}
To simplify the solution of this set, we use the matrix formalism. This allows us to express the system in terms of matrices, where $\mathbb{\Gamma}$ represents the total transfer matrix connecting region 1 to region 5. Thus, the set can be written as 
\begin{equation}\label{Max}
	\binom
	{\delta_{m,0}}
	{r_m}
	=\begin{pmatrix}
	\Gamma_{1,1}&	\Gamma_{1,2}\\
	\Gamma_{2,1}&	\Gamma_{2,2}
	\end{pmatrix}\binom
	{
	t_m}
	{\mathbb{0}_m}=\mathbb{\Gamma}\begin{pmatrix}
	t_m\\
	\mathbb{0}_m
	\end{pmatrix}
\end{equation}
where  $\mathbb{\Gamma}=\mathbb{\Gamma}(1,2)\cdot\mathbb{\Gamma}(2,3)\cdot\mathbb{\Gamma}(3,4)\cdot\mathbb{\Gamma}(4,5)$ and $\mathbb{\Gamma}(j,j+1)$ is a transfer matrix couples the region $j$ with region $j+1$, such as
\begin{align*}
	&\mathbb{\Gamma}(1,2)=\begin{pmatrix}
	\mathbb{I}&\mathbb{I}\\
		\mathbb{N}_1^+&\mathbb{N}_1^-
	\end{pmatrix}^{-1}\cdot \begin{pmatrix}
	\mathbb{J}_2&\mathbb{J}_2\\
		\mathbb{Q}_2^+&\mathbb{Q}_2^-
	\end{pmatrix},\quad
	\mathbb{\Gamma}(2,3)=\begin{pmatrix}
		\mathbb{E}^+&\mathbb{E}^-\\
		\mathbb{X}^+&\mathbb{X}^-
	\end{pmatrix}^{-1}\cdot \begin{pmatrix}
		\mathbb{Z}_1^+&\mathbb{Z}_1^-\\
	\mathbb{K}_1^+&\mathbb{K}_1^-
	\end{pmatrix}\\
	&\mathbb{\Gamma}(3,4)=\begin{pmatrix}
			\mathbb{Z}_2^+&\mathbb{Z}_2^-\\
		\mathbb{K}_2^+&\mathbb{K}_2^-
	\end{pmatrix}^{-1}\cdot \begin{pmatrix}
	\mathbb{J}_4&\mathbb{J}_4\\
	\mathbb{Q}_4^+&\mathbb{Q}_4^-
	\end{pmatrix},\quad
	\mathbb{\Gamma}(4,5)=\begin{pmatrix}
	\mathbb{S}^+&\mathbb{S}^-\\
	\mathbb{C}^+&\mathbb{C}^-
	\end{pmatrix}^{-1}\cdot\begin{pmatrix}
		\mathbb{I}&\mathbb{I}\\
		\mathbb{N}_5^+&\mathbb{N}_5^-
	\end{pmatrix}
\end{align*}
\begin{align*}
	&(\mathbb{N}^\pm_{1,5})_{m,l}=\pm(\Xi^m_{1,5})^{\pm 1}\delta_{m,l}, \quad
	(\mathbb{J}_2)_{m,l}=J_{m-l}(\alpha_2),\quad
	(\mathbb{Q}_2^\pm)_{m,l}=\pm \chi^{\pm 1}_2 J_{m-l}(\alpha_2),\quad
		(\mathbb{J}_4)_{m,l}=J_{m-l}(\alpha_4)e^{-i(m-l)\beta}\\
	&(\mathbb{Q}_4^\pm)_{m,l}=\pm \chi^{\pm 1}_2 J_{m-l}(\alpha_4)e^{-i(m-l)\beta},\quad
	(\mathbb{E}^\pm)_{m,l}=e^{\pm iq_2^ld}J_{m-l}(\alpha_2),\quad
		(\mathbb{X}^\pm)_{m,l}=\pm e^{\pm iq_2^ld}J_{m-l}(\alpha_2)\chi^{\pm 1}_2\\
		&(\mathbb{Z}_1^\pm)_{m,l}=\mu_m^\pm (-d_1)\delta_{m,l},\quad
		(\mathbb{Z}_2^\pm)_{m,l}=\mu_m^\pm (d_1)\delta_{m,l},\quad
			(\mathbb{K}_1^\pm)_{m,l}=\xi_m^\pm (-d_1)\delta_{m,l},\quad
		(\mathbb{K}_2^\pm)_{m,l}=\xi_m^\pm (d_1)\delta_{m,l}\\
		&(\mathbb{S}^\pm)_{m,l}=e^{\pm iq_4^ld}J_{m-l}(\alpha_4)e^{-i(m-l)\beta},\quad
		(\mathbb{C}^\pm)_{m,l}=\pm e^{\pm iq_4^ld}J_{m-l}(\alpha_4)\chi^\pm_4e^{-i(m-l)\beta}, \quad d=d_2-d_1.
\end{align*}
We can clearly see from \eqref{Max} that the transmission coefficients are given by
\begin{equation}
	t_m=\Gamma_{1,1}^{-1}\cdot\delta_{m,0}.
\end{equation}
The infinite order matrix $\Gamma_{1,1}$ prevents direct digitization of the problem. To overcome this, we limit the series to a finite number of terms, ranging from $-N$ to $N$, where $N$ satisfies the condition $N > \max(\alpha_2, \alpha_4)$ \cite{timepot, timepot2}. Consequently, we can express the series as 
\begin{align}
t_{-N+k} = \Gamma^{-1}_{11}[k+1, N+1], \quad k = 0, 1, \cdots, 2N.
\end{align}
To explicitly calculate the transmission $T_l$ and reflection $R_l$ probabilities, we first consider the current density. By applying the continuity equation, we show that the incident, reflected, and transmitted current densities are expressed as follows
\begin{align}
&J_\text{inc}^0=v_F(\Lambda_0+\Lambda^*_0)\\
&J_\text{tra}^l=v_Ft^*_lt_l(\Lambda'_l+\Lambda'^*_l)\\
&J_\text{ref}^l=v_Fr^*_lr_l(\Lambda_l+\Lambda^*_l)
\end{align}
giving raise 
\begin{align}
&	T_l=\frac{|J_\text{tra}^l|}{|J_\text{inc}^0|}=\frac{\cos \theta^l}{\cos \theta^0}|t_l|^2
	\\	
	&R_l=\frac{|J_\text{ref}^l|}{|J_\text{inc}^0|}=\frac{\cos \theta^l}{\cos \theta^0}|r_l|^2
\end{align}
where the two angles take the forms
\begin{align}
\cos \theta^l=\frac{k_5^l}{\sqrt{\left(k^l_5\right)^2+\left(k_y+\frac{d_1}{l^2_B}\right)^2}},\quad \cos \theta^0=\frac{k_1^0}{\sqrt{\left(k^0_1\right)^2+\left(k_y-\frac{d_1}{l^2_B}\right)^2}}.
\end{align}
It is clear that the sum over all modes gives the total transmission
\begin{equation}
	T=\sum_{l}T_l.
\end{equation}

 Since the analysis of all transmission modes is a large task, we will limit ourselves to the first three modes. These are the transmissions with zero photon exchange, corresponding to the central energy band, $l=0$, while the first sidebands are associated with $l=\pm 1$. Thus, we can express the system as 
\begin{align}
	&t_{-1}=\Gamma_{1,1}^{-1}[1,2], \quad t_{0}=\Gamma_{1,1}^{-1}[2,2], \quad t_{1}=\Gamma_{1,1}^{-1}[3,2].
\end{align}
In the low-energy regime, the dominance of single-photon processes over multi-photon interactions justifies the aforementioned approximation. A numerical analysis will be performed to clarify the results and to outline the dynamics of the system. This will be accomplished by focusing on a specific subset of channels and exploring different configurations of the physical parameters.

The zero temperature conductance is defined as the average flux of fermions over the half Fermi surface \cite{conduct1, conduct2}. Alternatively, it can be expressed as the integration of the total transmission coefficient $T$ over $k_y$ \cite{Biswas2021}, which is given by
	\begin{equation}
		G=\frac{G_0}{2\pi}\int_{-k_y^{\text{max}}}^{k_y^{\text{max}}}T(E,k_y)dk_y
	\end{equation}
	where $G_0$ represents the unit of conductance.
	Expressing $G$ in terms of the relationship between the transverse wave vector $k_y$ and the incident angle $\phi_0$
	\begin{equation}
		G=\frac{G_0}{2\pi}\int_{-\phi_0^{\text{max}}}^{\phi_0^{\text{max}}}T(E,E\sin\phi_0) \cos\phi_0d\phi_0
	\end{equation}
	where $\frac{\pi}{2}$ can be obtained from the relation $k_y^{\text{max}}=E\sin(\phi_0^{\text{max}})$.	In order to explore and highlight the fundamental characteristics of the current system, we perform a numerical analysis of the transport properties, focusing on the transmission channels and their associated conductance.

\section{Nemurical results}\label{NR}
\begin{figure}[ht]
	\centering
	\subfloat[]{\centering\includegraphics[scale=0.47]{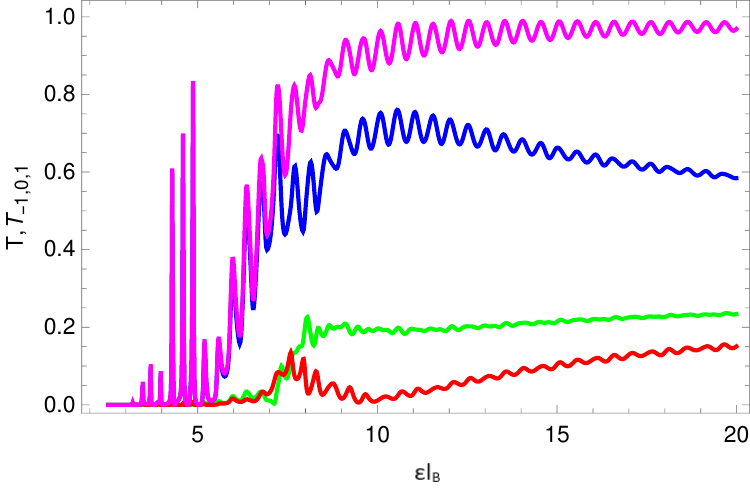}\label{fig3a}}
	\subfloat[]{\centering\includegraphics[scale=0.47]{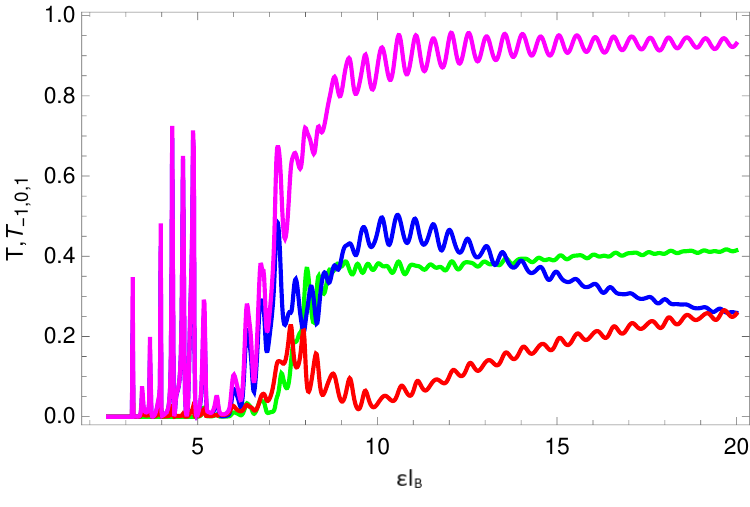}\label{fig3b}}
	\subfloat[]{\centering\includegraphics[scale=0.47]{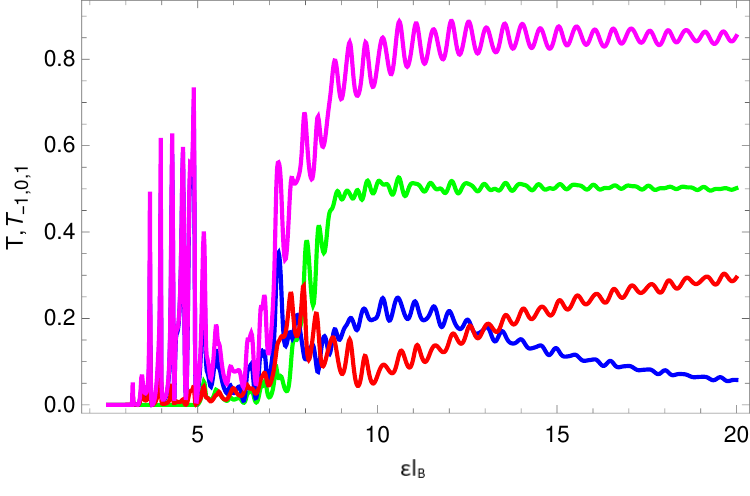}\label{fig3c}}
	\caption{The transmission probability as a function of incident energy $\varepsilon l_B$ for different values of laser field amplitudes $F_2 l_B=F_4 l_B=F$, for $k_y l_B  =0$, $d_1=3l_B$, $d_2=5l_B$, $\varpi l_B=1.8$ and $\beta=0$. (a): $F l_B=0.1$, (b): $F l_B=0.75$, (c):  $F l_B=2$, with $T$ (magenta line), $T_0$ (blue line), $T_1$ (red line), $T_{-1}$ (green line).}\label{fig3}
\end{figure}

In this section, we present our numerical results as a function of the system parameters, highlighting how different variables affect the behavior of the system.
 Fig. \ref{fig3} shows the transmission as a function of the incident energy of normally incident fermions for different values of the laser field amplitude. We have to check the following condition ($\varepsilon l_B>k_yl_B+\frac{d_1}{l_B}+l\varpi l_B$) to have transmission, this quantity plays the role of an effective mass \cite{masse}. Fig. \ref{fig3a}, plotted for a very low laser field ($F l_B=0.1$), shows that the transmissions vary in an oscillatory way with a dominance of the transmission without photon exchange. The total transmission oscillates around unity starting from $\varepsilon l_B=8$, which implies the appearance of the Klein tunneling effect, one of the main problems delaying the use of graphene in electronics. The transmission with photon absorption or emission is very low, not exceeding $20\%$, even when the energy is increased. When we increase the laser field intensity to $F l_B=0.75$ (see Fig. \ref{fig3b}), we observe that the total transmission still oscillates around unity, but the transmission with photon exchange becomes more significant than the transmission with zero photon exchange. 
 In Fig. \ref{fig3c}, as we increase the laser field intensity to $F l_B=2$, the total transmission decreases and does not exceed $90\%$. In this case, the lateral transmission is more significant than the central band transmission. Starting from $\varepsilon l_B=13$, the transmission with zero photon exchange becomes zero, i.e. all transmitted fermions cross the barrier with photon exchange. We can conclude that increasing the laser field decreases the total transmission, which helps to suppress the Klein tunneling effect, decreases the zero photon exchange transmission, and activates the process of transmission by photon absorption or emission, as we discussed in our previous work \cite{Elaitouni2023,Elaitouni2023A}.

\begin{figure}[ht]
		\centering
	\subfloat[]{\centering\includegraphics[scale=0.45]{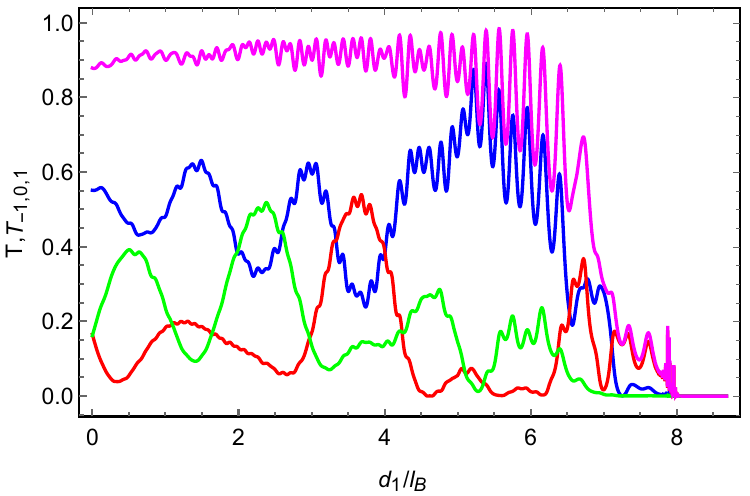}\label{fig4a}} 
	\subfloat[]{\centering\includegraphics[scale=0.45]{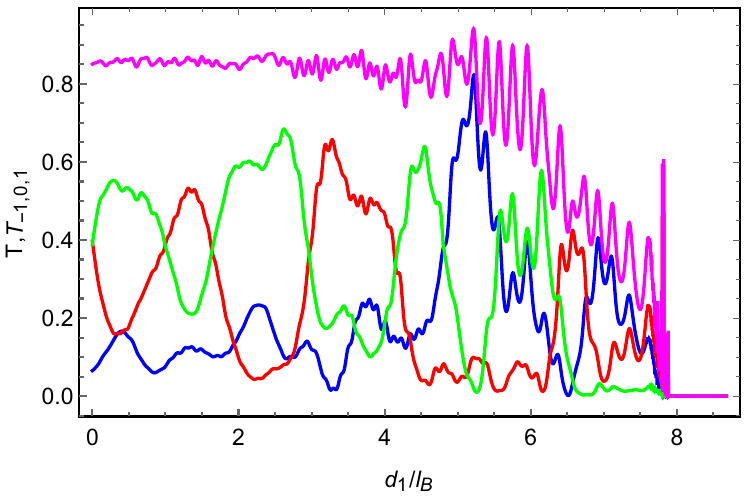}\label{fig4b}}
	\subfloat[]{\centering\includegraphics[scale=0.45]{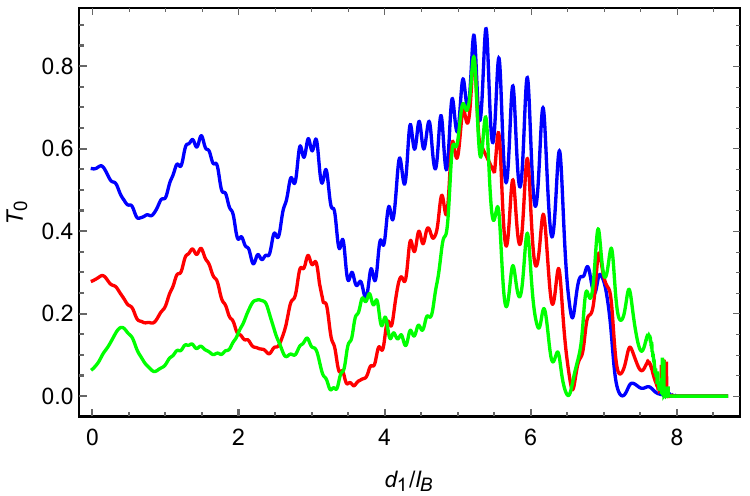}\label{fig4c}}
	
	\caption{The transmission as a function of $\frac{d_1}{l_B}$ for deffernets values of laser field amplitudes $F_2 l_B=F_4l_B=F$, for $k_yl_B =1$, $\frac{d_2}{l_B}=10$, $\varpi l_B=2$, $\varepsilon l_B=15$ and $\beta=0$. (a): $F=1.9$, (b): $F=3.9$, with $T$ (magenta line), $T_0$ (blue line), $T_1$ (red line), $T_{-1}$ (green line), (c): $F=1.9$ (blue line), $F=2.9$ (red line) and $F=3.9$ (green line).}\label{fig4}
\end{figure}

Fig. \ref{fig4} present the total transmission and the first two transmission modes as a function of the distance $d_1/l_B$ for different values of the laser field amplitudes $F$. In the three figures, the transmission becomes zero as $d_1$ approaches $8 l_B$, which corresponds to a distance of $16 l_B$. In this case, all fermions are blocked, and this distance represents the maximum distance between the two barriers to achieve transmission.
 In Fig. \ref{fig4a}, we observe that the total transmission oscillates around its maximum and then decreases rapidly as $d_1/l_B$ approaches $d_2/l_B$. The transmission with zero photon exchange also varies in the same way as the total transmission and is more dominant. The photon exchange transmission exceeds $50\%$ for certain values of $d_1/l_B$. As $F$ increases, the photon exchange process becomes more dominant, as seen in Fig. \ref{fig4b}, but the total transmission is not perfect. In Fig. \ref{fig4c} we show the transmission with zero photon exchange, it is observed that increasing the laser field amplitude suppresses this mode of transmission and activates the photon exchange process. In this case, most of the fermions pass through the barrier with photon exchange.
The variation of the distance between the two barriers ($d_1/l_B$) is a tool for controlling and guiding the transmission, one can control the selection of transmission processes based on this distance. As $d_1/l_B$ approaches $d_2/l_B$, the two barriers become like peaks with increasing interference between the wave functions across the barriers, allowing a decrease in transmission with zero photon exchange.

\begin{figure}[ht]
		\centering
\subfloat[]{\centering\includegraphics[scale=0.45]{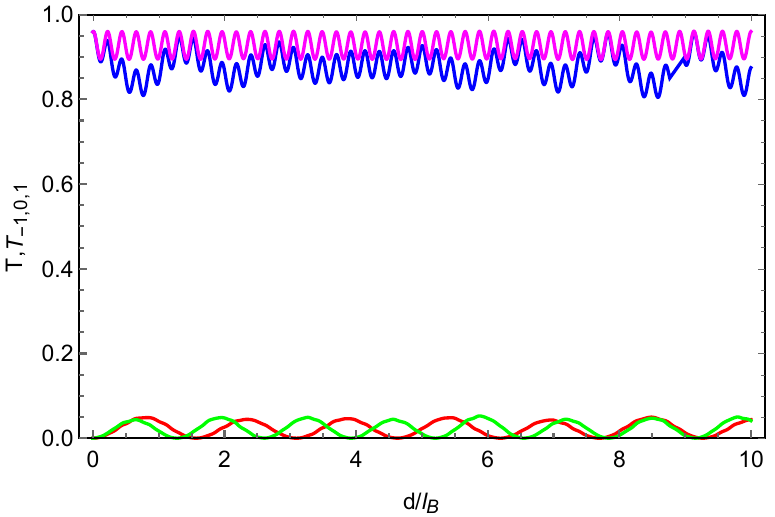}\label{fig5a}} \subfloat[]{\centering\includegraphics[scale=0.45]{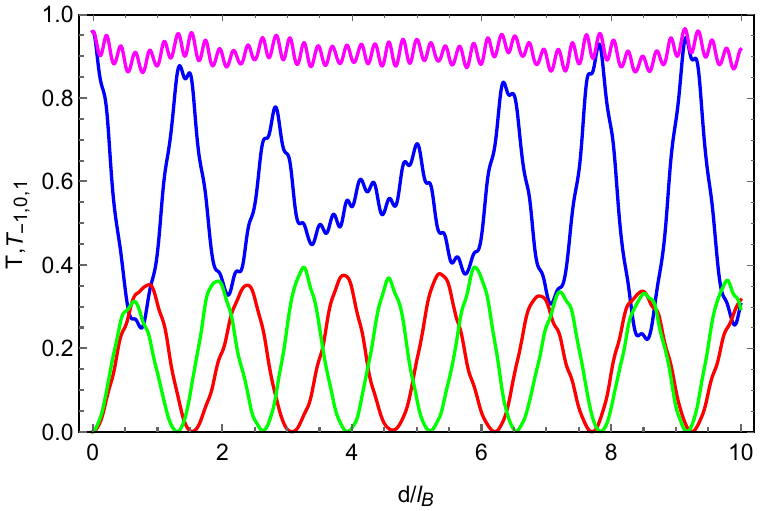}\label{fig5b}}
\subfloat[]{\centering\includegraphics[scale=0.45]{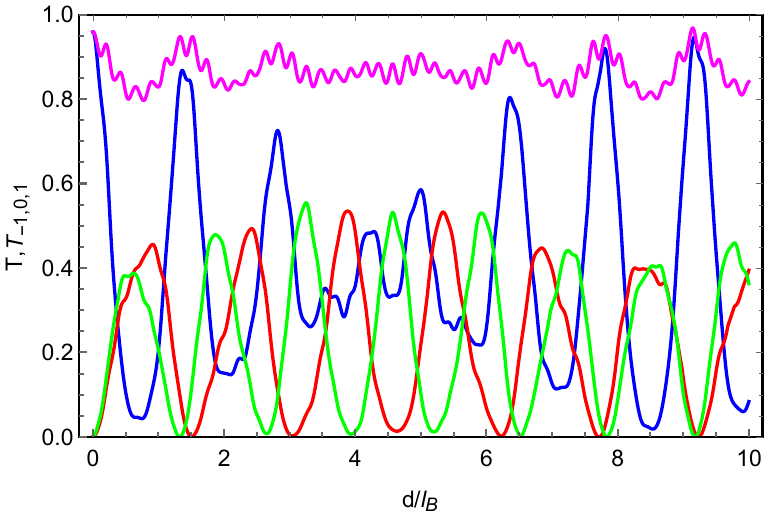}\label{fig5c}}
	\caption{The transmission as a function of the barrier width $\frac{d}{l_B}$ for different values of laser field amplitudes $F_2 l_B=F_4 l_B=F$, for $k_y l_B=0$, $\frac{d_1}{l_B}=3$, $\varepsilon l_B=15$, $\varpi l_B=2$ and $\beta=0$. (a): $F=0.9$, (b): $F=2.9$, (c): $F=3.9$, with $T$ (magenta line), $T_0$ (blue line), $T_1$ (red line), $T_{-1}$ (green line).}\label{fig5}
\end{figure}

In Fig. \ref{fig5}, we plot the transmission probability as a function of the barrier width $d/l_B$ for different values of $F$. Fig. \ref{fig5a} is plotted for $F=1.5$, and we observe that the total transmission oscillates around $1$, implying perfect transmission. We also observe that there are two transmission processes even at low field strengths. 
We see that the transmission process with zero photon exchange dominates and is almost equal to the total transmission, while the process with photon exchange varies oscillatorily around zero. For $F=2.9$ in Fig. \ref{fig5b}, we clearly observe the effect of the laser field on the transmission process. The transmission process with photon exchange becomes significant, exceeding $20\%$, and transmission with zero photon exchange oscillates.
However, we observe that the total transmission remains oscillating near unity. If we now increase the laser field intensity to $F=3.9$, i.e. $\alpha$ close to unity (Fig. \ref{fig5c}), the transmission through the central band oscillates from zero to $1$. We find points of transmission annihilation for certain values of $d/l_B$. We find that the transmission with photon absorption and emission exceeds $50\%$. In summary, in addition to controlling transmission by varying the laser field, we can also achieve control by adjusting the barrier width.

\begin{figure}[ht]
	\centering
	\subfloat[]{\centering\includegraphics[scale=0.45]{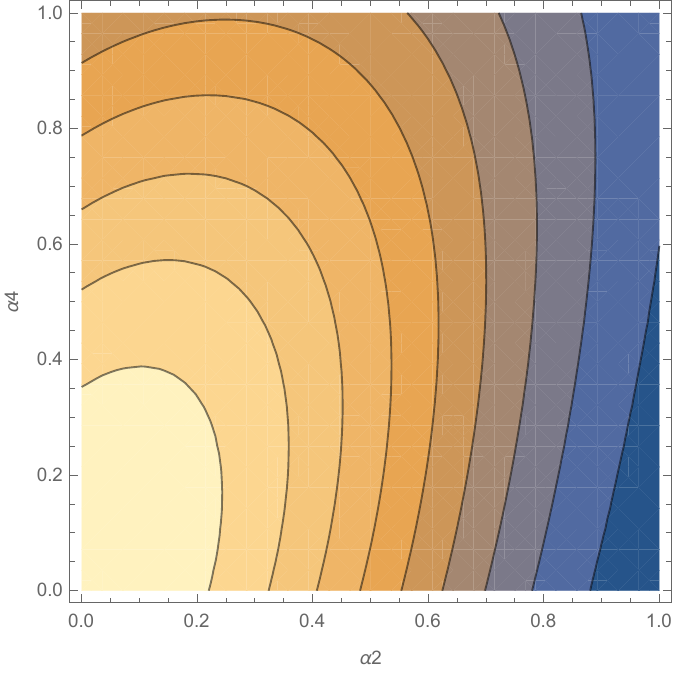}\label{fig6a}}
	 \subfloat[]{\centering\includegraphics[scale=0.45]{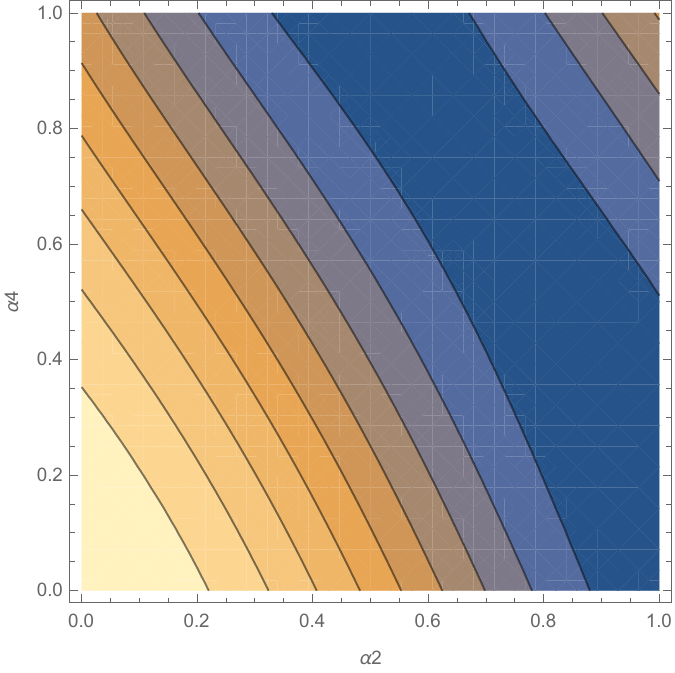}\label{fig6b}}
	\subfloat[]{\centering\includegraphics[scale=0.45]{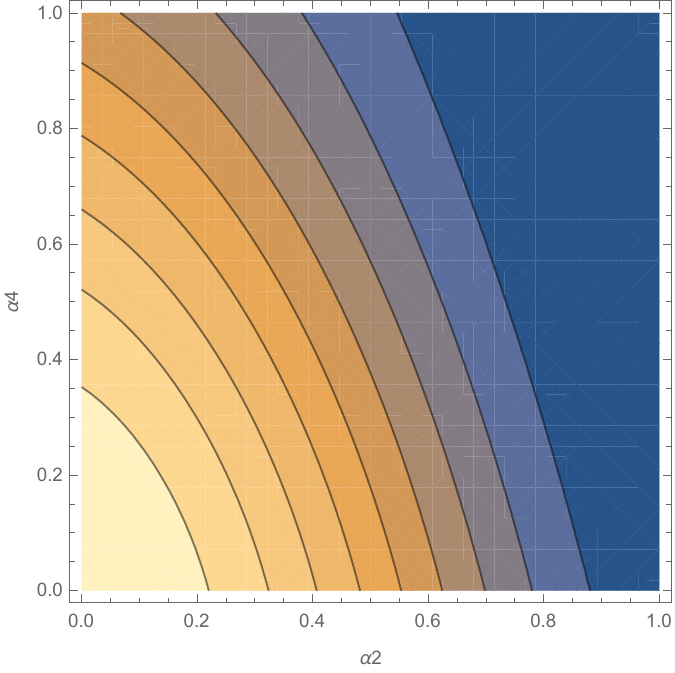}\label{fig6c}}
		\subfloat[]{\includegraphics[scale=0.45]{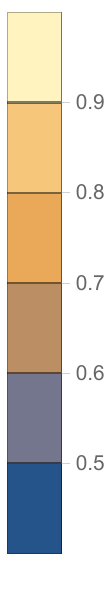}}
	\caption{Contourplot of transmission $T_0$ as a function of $\alpha_2=\frac{F_2}{\varpi^2}$ and $\alpha_4=\frac{F_4}{\varpi^2}$ for three value of $\beta$, for $k_y l_B=0$, $\frac{d_1}{l_B}=3$, $\frac{d_2}{l_B}=5$, $\varepsilon l_B=15$ and $\varpi l_B=2$. (a): $\beta=0$, (b): $\beta=\frac{\pi}{2}$, (c): $\beta=\pi $.}\label{fig6}
\end{figure}

Fig. \ref{fig6} is a contour plot of the transmission $T_0$ as a function of $\alpha_j=\frac{F_j}{\varpi^2} (j=2,4)$ for different values of the phase shift between the fields $\beta$. In Fig. \ref{fig6a}, where $\beta=0$, the two fields are synchronized, and the transmission $T_0$ is perfect only for the minimum values of $\alpha$. In this case, we have an additive effect of the two laser fields, which leads to a maximum transmission for the minimum values of $\alpha$.
In Fig. \ref{fig6b}, where the two fields are in phase opposition ($\beta=\frac{\pi}{2}$), the transmission is maximized for the lowest values of $\alpha$, with zones of equal probability appearing in parallel. In Fig. \ref{fig6c}, with $\beta=\pi$, the transmission is also maximized only for the minimum values of $\alpha$, and we observe that for $\alpha_2=1$, the transmission is minimal for all values of $\alpha_4$. In summary, the polarization of the laser fields plays a critical role in reducing the transmission through the central band and thereby increasing the transmission through the sidebands.

\begin{figure}[ht]
	
	\centering
	\subfloat[]{\centering\includegraphics[scale=0.45]{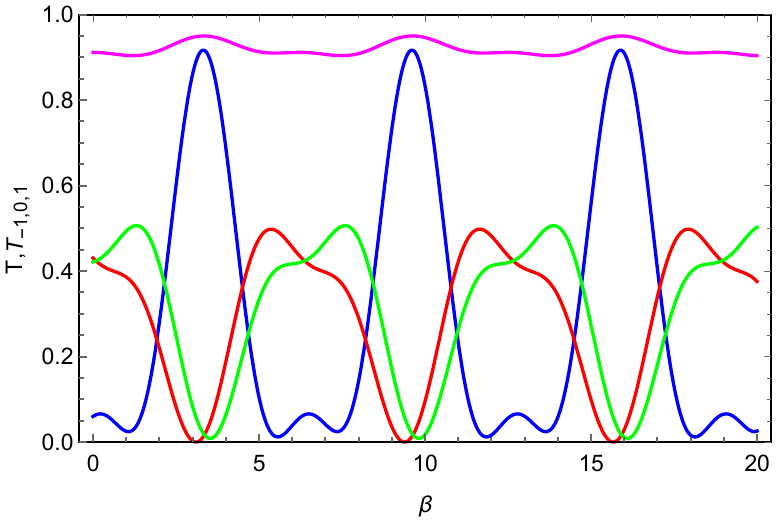}\label{fig2a}} \subfloat[]{\centering\includegraphics[scale=0.45]{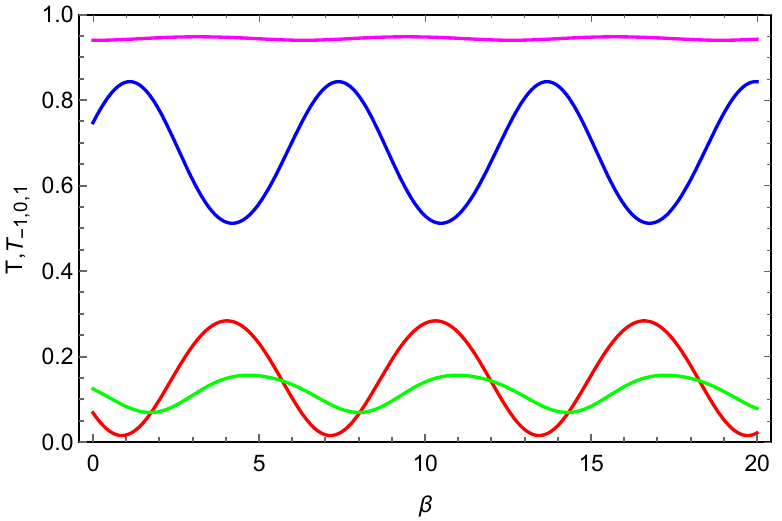}\label{fig2b}}
	\subfloat[]{\centering\includegraphics[scale=0.45]{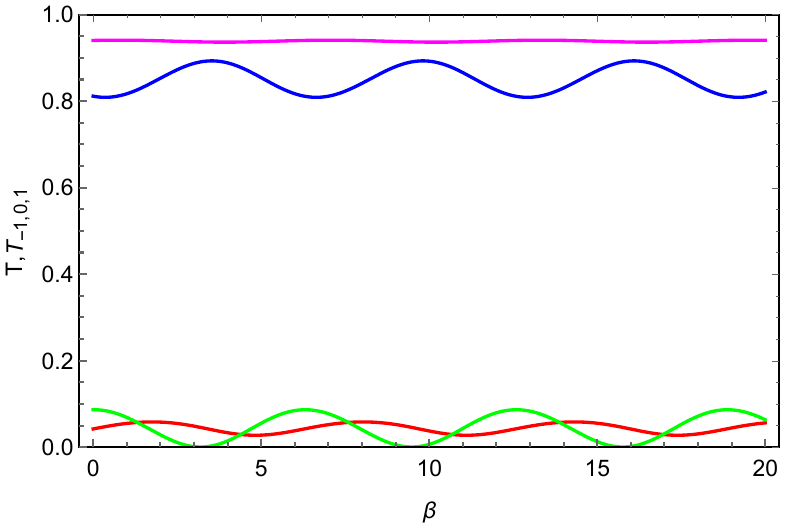}\label{fig2c}}
	\caption{The transmissions as a function of the phase shift $\beta$ between the two fields ($\beta$), for $d_1=3l_B$, $d_2=5l_B$, $F_2l_B=1.5$, $F_4 l_B=1.4$, $k_y l_B =1$ and $\varepsilon l_B=15$. (a): $\varpi l_B=1.5$, (b): $\varpi l_B=2$, (c): $\varpi l_B=2.5$, with $T$ (magenta line), $T_0$ (blue line), $T_1$ (red line), $T_{-1}$ (green line).}\label{fig2}
\end{figure}

Fig. \ref{fig2} shows the transmission as a function of the phase shift $\beta$ between the two laser fields for different values of $\varpi l_B$, $d_1=3l_B$, $d_2=5l_B$, $F_2l_B=1.5$, $\varepsilon l_B=15$, and $F_4l_B=1.4$. We observe that the transmission is not perfect for all values of $\beta$, indicating the suppression of the Klein tunneling effect. In Fig. \ref{fig2a} ($\varpi l_B=1.5$), the transmission with zero photon exchange is more dominant than the transmission process with photon exchange. The maximum transmission with zero photon exchange corresponds to the minimum transmission by the sidebands, with a periodicity of the zero points of both transmission processes. The transmission through the sidebands does not exceed $50\%$.
When the laser field frequency is increased to $\varpi l_B=2$ (see Fig. \ref{fig2b}), both transmission modes display oscillatory behavior, but the transmission without photon exchange is more dominant and does not vanish along $\beta$. Fig. \ref{fig2c}, plotted for $\varpi l_B=2.5$, clearly shows that the transmission process involving photon exchange weakens, while the transmission without photon exchange oscillates around 0.8. This indicates that most fermions cross the barrier without photon exchange. In summary, increasing the laser field frequency suppresses the transmission with photon exchange and enhances the transmission without photon exchange, as previously shown in our work \cite{doublelaser}. Similar results are observed for double barriers oscillating in time \cite{doubletemps}.

\begin{figure}[ht]	
	\centering
	\subfloat[]{\centering\includegraphics[scale=0.45]{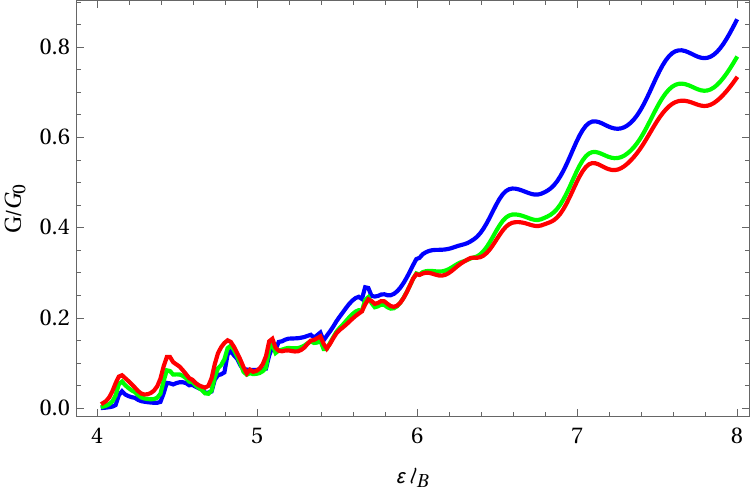}\label{fig7a}} \subfloat[]{\centering\includegraphics[scale=0.45]{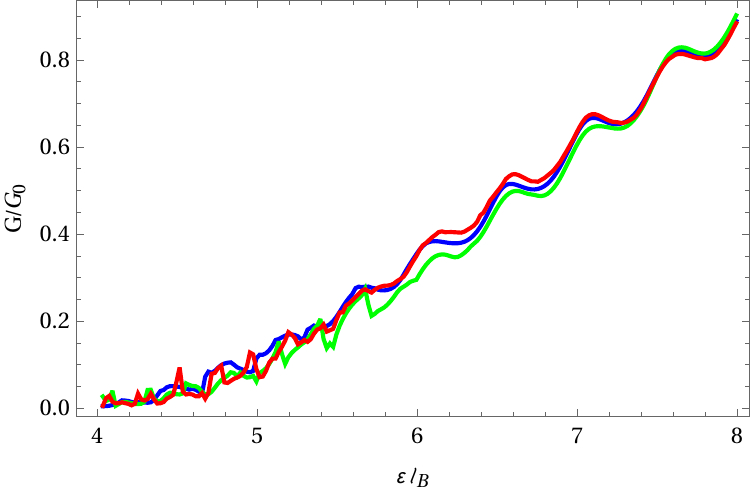}\label{fig7b}}
	\subfloat[]{\centering\includegraphics[scale=0.45]{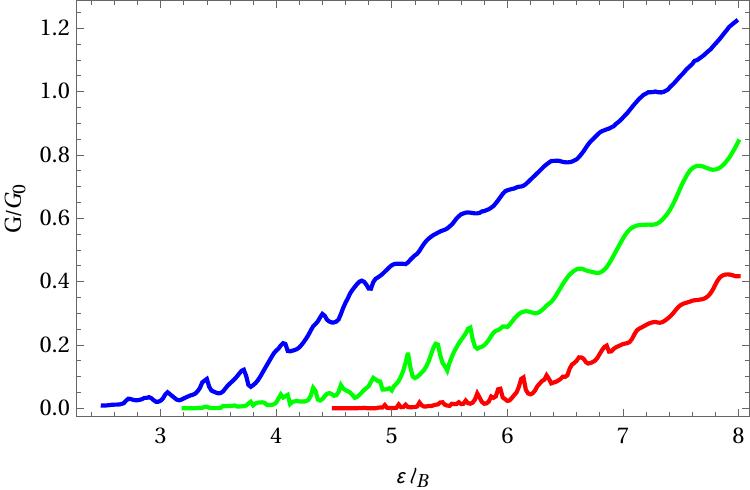}\label{fig7c}}
	\caption{The conductance $G/G_0$ as a function of incident energy $\varepsilon l_B$ for $\beta=0$, $F_1l_B=F_2l_B=F$ and $\varpi l_B=2$. (a): $d_1=3l_B$, $d2=5l_B$, $F=1.9$ (blue line), $F=2.9$ (green line) and $F=3.9$ (red line). (b): $F=2$, $d_1=3l_B$, $d_2=4l_B$ (blue line), $d_2=6l_B$ (green line) and $d_2=8l_B$ (red line). (c): $d_2= 6l_B$, $F=2.9$, $d_1=2l_B$ (blue line), $d_1=3l_B$ (green line), $d_1=4l_B$ (red line).}\label{fig7}
\end{figure}

Fig. \ref{fig7}  shows the conductance $G/G_0$ as a function of the incident energy $\varepsilon l_B$ by fixing the two parameters to $\beta=0$ and $\varpi l_B=2$. As a preliminary observation, we note that the conductance always increases regardless of the chosen configuration of the physical parameters.
In Fig. \ref{fig7a} we set the distance between the barriers to \(2d_1 = 6l_B\), the width of each barrier to \(d = d_2 - d_1 = 2l_B\), and both fields have the same intensity, varying as \(Fl_B = 1.9, 2.9, 3.9\). We observe that the conductance is inversely proportional to \(Fl_B\), since \(G/G_0\) decreases as \(Fl_B\) increases. This is consistent with the transmission behavior, where increasing \(Fl_B\) suppresses the tunneling effect. The case of varying $d_2$ and keeping $d_1=3l_B$ for $Fl_B=2.9$ is shown in Fig. \ref{fig7b}. It can be seen that \(G/G_0\) always increases but remains almost unchanged with the variation of \(d_2\). This happens because the transmissions are more easily channeled while the total transmission remains constant due to the change in barrier width. Therefore, even though the distance between the barriers varies, the conductance does not. 
The conductance for various inter-barrier distance values is shown in Fig. \ref{fig7c}. We find that as the inter-barrier distance increases, the Klein tunneling effect is suppressed and transmission is reduced, resulting in a decrease in conductance.

Finally, to illustrate the relevance and novelty of our results, we make the following remarks. In the case of a simple electrostatic barrier, the passage of fermions is independent of the height and width of the barrier. This means that the probability of fermions passing through the barrier is perfect for normal incidences, which is known as the Klein tunneling effect \cite{klien1}. However, the scenario changes significantly when dealing with magnetic barriers and laser-assisted barriers. In these cases, the transmission is influenced by all the parameters of the barrier, including its height, width, and the characteristics of the magnetic or laser fields. When the barrier oscillates in time \cite{timepot}, two different transmission processes occur. The first process involves photon exchange between the barrier and the fermions, where photons are absorbed or emitted during transmission. The second process occurs without any photon exchange. 
	It has been observed that the transmission without photon exchange is generally more significant than the photon exchange process \cite{bis2}. For a static barrier irradiated by a laser field, the transmission without photon exchange decreases proportionally with the amplitude of the laser field. At the same time, the transmission process with photon exchange becomes more pronounced. For a double barrier oscillating in time \cite{doubletemps}, both processes are always present, and the transmission can be channeled for certain values of the inter-barrier distance. The same is true for two laser barriers \cite{doublelaser}. If the inter-barrier region is controlled by a magnetic field, the transmission characteristics become even more complex. In this scenario, the transmission depends not only on the parameters of the laser fields, such as their intensity and frequency, but also on the intensity of the magnetic field. This interaction between the laser and magnetic fields can lead to a variety of transmission behaviors.

\section{Conclusion}\label{CC}
We studied the transmission through a structure of two laser fields separated by a region under the influence of a magnetic field. The periodicity of the lasers in time led us to use the Floquet approximation. The continuity of the wave function over the five regions yields eight equations, each with an infinite number of modes due to the infinite number of subbands created by the lasers. To simplify the calculation, we use matrix formalism. The continuity equation allows the calculation of the current density for each region, which in turn allows the calculation of the transmission corresponding to each energy band and the derivation of the total transmission. 

The time-dependent oscillation of the barrier leads to a degeneration of the energy spectrum since the barrier has an energy level at each instant. This quantization gives rise to two transmission processes: transmission with zero photon exchange between the barrier and the fermions, corresponding to the central band, and transmission with photon exchange, corresponding to the side bands. The challenge of digitizing all modes with our machine limits our study to the first side band corresponding to $l=0,\pm1$. Transmission exists if the incident energy satisfies the following condition ($\varepsilon l_B> k_yl_B+\frac{d_1}{l_B}+l\varpi l_B$), so all fermions with energies that do not satisfy this condition are blocked. The transmission through the central band is more dominant than that through the sidebands, implying that the majority of fermions cross the barrier with zero photon exchange.
 However, increasing the amplitude of the laser field can suppress the zero-photon transmission process and activate the photon exchange process. Increasing the laser irradiation frequency has an inverse effect on transmission. Enlarging the region controlled by the magnetic field allows to reduce the transmission in all bands, which implies a decrease in conductance. The polarization of the laser fields plays a crucial role in controlling the passage of fermions and guiding the transmission channel. By tuning the polarization, we can control the transmission through the barrier. These results provide opportunities for the design of electronic devices based on graphene.
 
 We have found that the conductance is affected by several factors, including the variation of the laser field intensity and the distance between the barriers where the magnetic field is applied. When the laser field intensity varies, it changes the energy of the electrons, which can affect their ability to cross the barriers. Similarly, an increase in the inter-barrier distance where the magnetic field is applied can increase the interactions of the electrons with the magnetic field, thereby decreasing their transmission and consequently the conductance. 
The laser fields play a crucial role in controlling the passage of the fermions and the direction of the transmission channel. By adjusting the polarization, we can not only influence the trajectory of the fermions, but also modulate the efficiency and selectivity of the transmission channel. This ability to control transmission through the barrier allows the optimization of particle filtering and guiding processes, paving the way for advanced applications in quantum physics and communication technologies. In summary, precise manipulation of laser field polarization is a powerful tool for controlling interactions at the microscopic scale.


\begin{thebibliography}{}
	\bibitem{disc}
K. S. Novoselov, A. K. Geim, S. V. Morozov, D. Jiang, Y. Zhang, S. V. Dubonos, I. V. Grigorieva, and A. A. Firsov, Science 306, 666 (2004).
\bibitem{prop}
A. H. Castro Neto, F. Guinea, N. M. R. Peres, K. S. Novoselov, and A. K. Geim, Rev. Mod. Phys. 81, 109 (2009). 
\bibitem{mobile}
K. I. Bolotin, K. J. Sikes, Z. Jiang, M. Klima, G. Fudenberg, J. Hone, P. Kim, and H. L. Stormer, Solid State Commun. 351. 146 (2008).
\bibitem{masless}
A. K. Geim, Science 324, 1530 (2009).
\bibitem{mobil2}
S. V. Morozov, K. S. Novoselov, M. I. Katsnelson, F. Schedin, D. C. Elias, J. A. Jaszczak, and A. K. Geim, Phys. Rev. Lett. 100, 016602
(2008).

\bibitem{absor}Qiaoliang Bao, Han Zhang, Bing Wang, Zhenhua Ni, Candy Haley Yi Xuan Lim, Yu Wang, Ding Yuan Tang, and Kian Ping Loh,
Nature Photon 5, 411 (2011).
\bibitem{sp2}
Pierre Petit and Annick Loiseau, Comptes Rendus. Physique 4, 967 (2003).
\bibitem{zero}
G. Gui, J. Li, and J. Zhong, Phys. Rev. B 78, 075435 (2008).
\bibitem{desp2}
N. M. R. Peres, J. Phys.: Condens. Matter 21, 323201 (2009).
\bibitem{zero1}
Y. Zheng and T. Ando, Phys. Rev. B 65, 245420 (2002).
\bibitem{Tight}
S. Reich, J. Maultzsch, C. Thomsen, and P. Ordejon, Phys. Rev. B 66, 035412 (2002).


\bibitem{klien1}
M. I. Katsnelson, K. S. Novoselov, and A. K. Geim, Nat. Phys. 2, 620 (2006).
\bibitem{klien2}
C. W. J. Beenakker, Rev. Mod. Phys. 80, 1337 (2008). 
\bibitem{klienexp}
N. Stander, B. Huard, and D. Goldhaber-Gordon, Phys. Rev. Lett. 102, 026807 (2009).

\bibitem{dopage} G. Giovannetti, P. A. Khomyakov, G. Brocks, V. M. Karpan, J. van den Brink, and P. J. Kelly, Phys. Rev. Lett. 101, 026803 (2008).
\bibitem{def1}
F. Guinea, M. I. Katsnelson, and A. K. Geim, Nat. Phys. 6, 30 (2010).
\bibitem{def2}
G.-X. Ni, Y. Zheng, S. Bae, H. R. Kim, A. Pachoud, Y. S. Kim, C.-L. Tan, D. Im, J.-H. Ahn, B. H. Hong, and B. Ozyilmaz, ACS Nano 6, 1158 (2012).
\bibitem{substrat}
A. N. Sidorov, M. M. Yazdanpanah, R. Jalilian, P. J. Ouseph, R. W. Cohn,  and G. U. Sumanasekera, Nanotechnology 18, 135301 (2007).
\bibitem{mag1}
L. Dell'Anna and A. De Martino, Phys. Rev. B 79, 045420 (2009).
\bibitem{mag4}
A. Jellal and A. El Mouhafid, J. Phys. A: Math. Theo. 44, 015302 (2011).
\bibitem{Elaitouni2023}
R. El Aitouni, M. Mekkaoui, A. Jellal, and M. Schreiber, Phys. E 157, 115865 (2023).
\bibitem{Elaitouni2023A}
R. El Aitouni, M. Mekkaoui, and A. Jellal, Ann. Phys. (Berlin) 535, 2200630 (2023).
\bibitem{mag3}
M. R. Masir,  P. Vasilopoulos, and F. M. Peeters, Phys. Rev. B 77, 235443 (2008).
\bibitem{oscil1}
A. Jellal, M. Mekkaoui, E. B. Choubabi, and H. Bahlouli, Eur. Phys. J. B 87, 123 (2014).
\bibitem{oscil2}
C. Zhang and N. Tzoar, Appl. Phys. Lett. 53, 1982 (1988).
\bibitem{doublelaser}
R. El Aitouni, M. Mekkaoui,  and A. Jellal, Phys. Scr. 99, 065912 (2024). 
\bibitem{bis2}
R. Biswas and C. Sinha, Appl. Phys. 114, 183706 (2013).
	\bibitem{dipole}
R. Loudon, The Quantum Theory of Light, 3rd ed. (Oxford University
Press Inc., New York, 2000).

\bibitem{bessel}
	Z. Li. J. Wu and K. S. Chan, Phys. Lett. A, 376, 1159 (2012).
	\bibitem{floq}
	J. H. Shirley, Phys. Rev. 138, B979 (1965).

\bibitem{timepot}
M. Ahsan Zeb, K. Sabeeh, and M. Tahir, Phys. Rev. B 78, 165420 (2008).
\bibitem{timepot2}
Wenjun Li and L. E. Reichl, Phys. Rev. B 60, 15732 (1999).
\bibitem{doubletemps}
H. P. Ojeda-Collado and C. Rodríguez-Castellanos, Appl. Phys. Lett. 103, 033110 (2013).
\bibitem{Elaitouni2022}
R. El Aitouni and A. Jellal, Phys. Lett. A 447, 128288 (2022).
\bibitem{magneticfield}
S. Park and H. S. Sim, Phys. Rev. B 77, 075433 (2008).

\bibitem{masse}
M. Mekkaoui, A. Jellal, and H. Bahlouli, Solid State Communi. 358, 114981 (2022).

	
	\bibitem{conduct2}
	M. R. Masir, P. Vasilopoulos, and F. M. Peeters, Phys.
	Rev. B 79, 035409 (2009).
	
	\bibitem{conduct1}
	X. Chen and J. W. Tao, Appl. Phys. Lett. 94, 262102 (2009).
	
	
	
	\bibitem{Biswas2021}
	R. Biswas and C. Sinha, Sci. Rep. 11, 2881 (2021).




\end{thebibliography}
\end{document}